\documentstyle[prc,preprint,aps]{revtex}
\draft
\begin{document}
\def\ii{\'\i}
\title{Transition from the Seniority to the Anharmonic Vibrator
Regime in Nuclei}
\author{R. Bijker$^1$, A. Frank$^{1,2}$ and S. Pittel$^3$} 
\address{$^1$Instituto de Ciencias Nucleares, U.N.A.M., 
A.P. 70-543, 04510 M\'{e}xico, D. F., M\'{e}xico \\ 
$^2$Instituto de F\ii sica, Laboratorio de Cuernavaca, U.N.A.M., 
A.P. 139-B, Cuernavaca, Morelos, M\'{e}xico \\ 
$^3$Bartol Research Institute, University of Delaware, 
Newark, DE 19716, U.S.A. }
\maketitle

\begin{abstract}
A recent analysis of experimental energy systematics suggests that all 
collective nuclei fall into one of three classes -- seniority, 
anharmonic vibrational, or rotational -- with sharp phase transitions 
between them.  We investigate the transition from the seniority 
to the anharmonic vibrator regime 
within a shell model framework involving a single large $j$--orbit.
The calculations qualitatively reproduce the observed transitional behavior, 
both for $U(5)$ like and $O(6)$ like nuclei. They also confirm the preeminent 
role played by the neutron-proton interaction in producing the phase 
transition.

\end{abstract} 

\begin{center}
\end{center}

\clearpage

A recent analysis of the energy systematics of the lowest 2$^+$ and 4$^+$ 
states
in collective even--even nuclei has shown some remarkable 
features\cite{r:Casten1,r:Casten2}. 
These data suggest
that medium--weight and heavy collective even-even nuclei 
fall into three classes: (a) nuclei dominated by a generalized--seniority
structure, (b) anharmonic vibrational nuclei, and 
(c) well--deformed rotational nuclei.
In each of these three regimes, the energy systematics is extremely robust,
suggesting an underlying simplicity of low--energy nuclear structure never
before appreciated. Furthermore, the transitions between regimes occur very
rapidly, typically with the addition or removal of only one or two pairs
of nucleons. Since these observations were reported, 
much work has been carried out
to understand their significance. There is now a fairly good -- albeit
general -- understanding of the three regimes themselves \cite{r:Casten2} and
of the transition from anharmonic vibrational to rotational 
nuclei. 

In this work we study the transition from the seniority to the anharmonic 
vibrator regime. In the seniority region, the energies of the lowest 
4$^+$ and 2$^+$ states are related by the simple formula 
$E(4_1^+)~=~ E(2_1^+)~+~\epsilon_4$, where $\epsilon_4$ is
constant in each nuclear region. Nuclei falling into this class are typically
either semi-magic or nearly semi-magic. For the anharmonic vibrator regime, 
on the other hand, the corresponding relation is 
$E(4_1^+)~=~ 2E(2_1^+)~+~\epsilon_4$, with
$\epsilon_4 \approx 156$ keV in the entire $Z=38-82$ region 
and with $\epsilon_4 \approx 67$ keV in the actinides. 
Nuclei that fall into this class include those characteristically 
thought of as vibrational (or $U(5)$ like)
and gamma-soft (or $O(6)$ like). 

To gain a more fundamental insight into this transition,
we have carried out a series of shell-model calculations. The
calculations were performed under the simplifying assumption that the 
neutrons and protons each occupy a single large $j=15/2$ orbital. 
The calculations were carried out for 0 to 6 protons with either
0 to 6 neutrons or 0 to 6 neutron holes, since in both cases the 
data show transitions to an anharmonic vibrational behavior. 
  
Despite the assumption of a single active $j$--orbit,
the dimensions increase rapidly, reaching 3199 for four
active protons ($Z=4$), four active neutrons ($N=4$)
and $J^{\pi}=4^+$. 
Increasing the angular momentum of the single $j$--orbit or
increasing the number of active particles make the calculations
prohibitive. As such, the calculations that we report should 
not be directly interpreted in terms of a real nuclear region, but rather
as a schematic study of the mechanism underlying the transition 
from one regime to the other. Similarly, the hamiltonian strength 
parameters, described below, 
represent effective values. 

We use a shell-model hamiltonian with the following ingredients:
\begin{itemize}
\item
The interaction
between protons is assumed to be a surface delta interaction (SDI) of 
strength $G_{pp}=0.24$ MeV,
supplemented by a quadrupole interaction of strength 
$\chi_{pp}=0.6$ MeV to bring down the energy of the lowest
$2^+$ state in semi-magic nuclei to a value compatible with data for
such nuclei. 
\item
The interaction between neutrons is likewise taken to be
a quadrupole-enhanced SDI. When the neutrons are particle--like,
we choose a strength $G_{nn}=0.22$ MeV, somewhat weaker than for protons to 
simulate a small neutron excess. When they are hole--like, so that 
the excess is even larger, we assume a strength of
$G_{nn}=0.20~MeV$. In both cases, the strength of the quadrupole 
interaction is chosen to be $\chi_{nn}=1.0$ MeV. 
\item
Finally, for the neutron--proton interaction, we 
assume a pure quadrupole force, with strength $\chi_{np}=10$ MeV. 
\end{itemize}
The quadrupole strength parameters given above include the single relevant
radial matrix element.

Table \ref{t:energies} summarizes the results we obtain for the 
energies of the lowest
$2^+$ and $4^+$ states for all systems that we considered. 
The results clearly exhibit the systematics of the
data noted earlier. Namely:
\begin{itemize}
\item
For semi--magic nuclei, the energy of the lowest $4^+$ state in all cases is
a few hundred keV above that of the lowest $2^+$ state. This is a
pattern typical of spherical nuclei in the seniority regime. 
\item
When particles are either added to or removed from the closed shell of the
semi--magic nucleus, there is a
rapid change in the behavior of the lowest $2^+$ and $4^+$ energies. When
both the neutrons and protons are particle--like, adding just two particles
to a semi--magic seniority nucleus raises the ratio $R=E(4_1^+)/E(2_1^+)$ to
a value greater than 2 (except when $Z=N=2$ where it only rises to
1.8). Successive addition of nucleons further raises this ratio, albeit only
slightly. A similar, though not quite as strong, effect is seen for
nuclei involving proton particles and neutron holes. Adding two holes
to the semi--magic nucleus quickly raises $R$ 
to about 1.8 or 1.9 (except when $Z=2$ and $N=-2$ where it is smaller).
This is precisely the pattern seen in the data; a rapid transition from
the seniority regime to the anharmonic vibrator regime. 
Especially interesting
is that (as noted above) the same behavior was found 
(with minor quantitative differences) in the
transitions to $U(5)$ nuclei (both neutrons and protons particle--like)
and $O(6)$ nuclei (one type of nucleon particle--like and the other 
hole--like).
\end{itemize}

The fact that our calculations, despite their simplicity, 
reproduce reasonably well the qualitative trends observed in the energy 
systematics lends hope that they may shed light on the physics
underlying the phase transition. With that in mind, we now turn to
a discussion of the wave functions that emerged from our calculations.
We concentrate on a limited set of representative results to illustrate 
the key points. 

In Table \ref{t:wavefunctions}, we present results for systems with 4 
active protons and
differing numbers of active neutrons or neutron holes ranging from 0 to 4.
More specifically, we give the probability amplitudes for different seniority
configurations ($v_p,v_n$), where $v_p$ and $v_n$ denote the
proton and neutron seniorities, respectively. (Note that since we have
limited our analysis to a single $j$--orbit, these are real and not 
generalized seniorities \cite{Talmi}.)

When there is only one type of nucleon ($Z=4,~N=0$), the lowest $2^+$ and
$4^+$ states are almost pure in their seniority structure. That they are not
exact seniority eigenstates is a consequence of the
additional quadrupole component included in the two-proton interaction. 

As neutron pairs are added or removed, the structure of the eigenstates 
changes 
dramatically. The lowest $2^+$ and $4^+$ states immediately develop
a strongly mixed seniority structure, due to the seniority--breaking 
neutron--proton
quadrupole force. The mixing is already significant with the addition of only
2 neutrons (or neutron holes) and becomes stronger with the addition of 
two more. Associated directly with this seniority mixing is the rapid 
development
of an anharmonic vibrator energy pattern.

That the neutron--proton interaction is responsible for the onset of 
collectivity and that seniority mixing plays a central role in this
phenomenon are already well known facts\cite{r:PF}. There are two 
specific features
that appear in our results, however, 
which we do not believe are as well known and on which we
therefore wish to elaborate.
The first is that precisely the same features occur when one
adds neutrons or neutron holes. From the table, we see that qualitatively the
same mixing shows up in both cases. Quantitatively, the mixing is somewhat 
weaker in the
case of neutron holes, and this is directly correlated with the fact that
the $4^+$ to $2^+$ energy ratio is not quite as large there. 

The second point we would like to note concerns some important differences 
that evolve in the structure of the lowest $2^+$ and $4^+$ states
when neutrons or neutron holes are added to the semi--magic system.
As we can see from the table, the  $4^+$
wave functions typically have a larger component of 
$(v_p,v_n)~=~(4,0)$ than
$(2,0)$. This is to be contrasted with the $2^+$ states, for which
the $(2,0)$ configuration is always more important. What is the reason for this
and what (if anything) is its impact on nuclear collectivity?

The answer to these questions arises from considering the pair structure
of these seniority configurations. In the case of $2^+$ states, the
$(2,0)$ configuration involves a single $D$--pair (with $J^{\pi}=2^+$), 
while the lowest $(4,0)$ configuration involves two $D$--pairs. 
In the case of $4^+$ states, the $(2,0)$ configuration involves a single 
$G$--pair (with $J^{\pi}=4^+$) and the lowest $(4,0)$ configuration 
corresponds to two $D$--pairs. 
In the semi--magic system, the $4^+$ configuration with a single $G$--pair 
is lower than that with two $D$--pairs (for any reasonable force).
What our results indicate is that
when particles of the other type are added to (or removed from) a semi--magic
nucleus, the configuration with two $D$--pairs is {\it effectively} lowered 
below that of the single $G$--pair configuration. 
A more detailed analysis shows
that this lowering takes place through off--diagonal (and not diagonal)
matrix elements of the  neutron--proton interaction, and this is the reason 
we use the term {\it effectively} to describe the phenomenon.
This effective crossing has important consequences on the 
description of collective properties of nuclei. It provides the mechanism
for $S$--$D$ pair dominance in the low--energy properties of nuclei 
with a large 
enough number of valence protons and neutrons, which is at the heart of
the success of the  Interacting Boson Model\cite{r:IBM}.

Summarizing, the calculations reported here qualitatively reproduce the
recently noted transition in $4^+-2^+$ energetics from seniority 
(or pre--collective) nuclei to anharmonic vibrational nuclei. 
{\it Energetically},
the behavior is no different when undergoing a transition to nuclei of 
$U(5)$ character or $O(6)$ character. In both cases, 
it is the neutron--proton interaction
which is responsible for the transition. The calculations also shed light on
how $S$--$D$ pair dominance emerges when there are enough active 
neutrons and active protons.

The authors would like to acknowledge useful conversations with Rick Casten
and Franco Iachello. This work was supported in part by the National Science
Foundation under grant \#s PHY-9303041, PHY-9600445 and INT-9314535, the
European Community through project CI1$^*$-CT94-0072, DGAPA UNAM through 
project IN105194 and CONACyT, Mexico, through project 400340-5-3401E.

\begin{table}
\caption{Calculated energies of the lowest $2^+$ and $4^+$ states for all 
nuclei considered in this study. $Z$ denotes the number of active protons
and $N$ the number of active neutrons (if positive) or neutron holes (if 
negative). $R$ denotes the ratio $R=E(4_1^+)/E(2_1^+)$.
All energies are given in MeV.
\label{t:energies}}
\vspace{15pt}
\begin{tabular}{ccccccc|cccccc} 
& & & & & & & & & & & & \\
& $Z$ & $N$ & $E_{2_1^+}$ & $E_{4_1^+}$ & $R$ & 
& $Z$ & $N$ & $E_{2_1^+}$ & $E_{4_1^+}$ & $R$ & \\
& & & & & & & & & & & & \\
\tableline
& & & & & & & & & & & & \\
& 0 & 2 & 1.35 & 1.61 & 1.19 && 0 & --2 & 1.23 & 1.47 & 1.19 & \\
& 0 & 4 & 1.27 & 1.70 & 1.34 && 0 & --4 & 1.15 & 1.56 & 1.36 & \\
& 0 & 6 & 1.23 & 1.76 & 1.42 && 0 & --6 & 1.11 & 1.62 & 1.45 & \\
& 2 & 0 & 1.46 & 1.71 & 1.17 && 2 & --0 & 1.46 & 1.71 & 1.17 & \\
& 2 & 2 & 0.56 & 1.01 & 1.79 && 2 & --2 & 0.83 & 1.23 & 1.49 & \\
& 2 & 4 & 0.38 & 0.82 & 2.17 && 2 & --4 & 0.64 & 1.11 & 1.74 & \\
& 2 & 6 & 0.35 & 0.79 & 2.24 && 2 & --6 & 0.49 & 0.94 & 1.93 & \\
& 4 & 0 & 1.41 & 1.77 & 1.25 && 4 & --0 & 1.41 & 1.77 & 1.25 & \\
& 4 & 2 & 0.40 & 0.85 & 2.14 && 4 & --2 & 0.65 & 1.11 & 1.72 & \\
& 4 & 4 & 0.23 & 0.59 & 2.51 && 4 & --4 & 0.48 & 0.92 & 1.93 & \\
& 6 & 0 & 1.39 & 1.80 & 1.30 && 6 & --0 & 1.39 & 1.80 & 1.30 & \\
& 6 & 2 & 0.38 & 0.84 & 2.21 && 6 & --2 & 0.51 & 0.98 & 1.94 & \\
& & & & & & & & & & & & \\
\end{tabular}
\end{table}

\clearpage 

\begin{table}
\caption{Probability amplitudes of seniority configurations $(v_p,v_n)$
for the lowest $2^+$ and $4^+$ states obtained in the calculations described
in the text.
\label{t:wavefunctions}}
\vspace{15pt}
\begin{tabular}{crrcc|ccccccccc}
& & & & & & & & & & & & & \\
& $Z$ & $N$ & $J^{\pi}$ && \multicolumn{8}{c}{$(v_p,v_n)$} & \\
&&&&& (2,0) & (4,0) & (0,2) & (0,4) & (2,2) & (2,4) & (4,2) & (4,4) & \\
& & & & & & & & & & & & & \\
\tableline
& & & & & & & & & & & & & \\
& 4 &   0 & 2$^+$ && 0.99 & 0.01 & --   & --   & --   & --   & --   & --   & \\
&   &     & 4$^+$ && 0.99 & 0.01 & --   & --   & --   & --   & --   & --   & \\
& 4 &   2 & 2$^+$ && 0.23 & 0.04 & 0.21 & --   & 0.25 & --   & 0.27 & --   & \\
&   &     & 4$^+$ && 0.03 & 0.15 & 0.14 & --   & 0.35 & --   & 0.33 & --   & \\
& 4 &   4 & 2$^+$ && 0.12 & 0.03 & 0.12 & 0.03 & 0.12 & 0.18 & 0.17 & 0.23 & \\
&   &     & 4$^+$ && 0.01 & 0.08 & 0.01 & 0.09 & 0.15 & 0.20 & 0.18 & 0.28 & \\
& 4 & --2 & 2$^+$ && 0.28 & 0.01 & 0.41 & --   & 0.20 & --   & 0.11 & --   & \\
&   &     & 4$^+$ && 0.08 & 0.06 & 0.37 & --   & 0.37 & --   & 0.12 & --   & \\
& 4 & --4 & 2$^+$ && 0.22 & 0.01 & 0.29 & 0.02 & 0.08 & 0.18 & 0.13 & 0.08 & \\
&   &     & 4$^+$ && 0.05 & 0.08 & 0.06 & 0.16 & 0.23 & 0.17 & 0.11 & 0.14 & \\
& & & & & & & & & & & & & \\
\end{tabular}
\end{table}

\end{document}